\documentclass[10pt,aip,cha,twocolumn,amsmath,showpacs,superscriptaddress,nofootinbib]{revtex4-1}

\usepackage[colorlinks,linkcolor=blue,citecolor=magenta,urlcolor=blue]{hyperref}
\usepackage{multirow}
\usepackage{graphicx}
\usepackage{txfonts}
\usepackage[all]{xy}

\def\CD{{\cal D}}
\def\CE{{\cal E}}

\def\CM{{\cal M}}
\def\CMini{{\CM_0}}
\def\rhoini{\rho^{(0)}}

\def\CS{{\cal S}}
\def\CI{{\cal I}}
\def\CV{{\cal V}}

\def\beq{\begin{eqnarray}}
\def\eeq{\end{eqnarray}}
\def\beqq{\begin{eqnarray*}}
\def\eeqq{\end{eqnarray*}}

\def\be{\begin{equation}}
\def\ee{\end{equation}}

\def\lb{\left}
\def\rb{\right}

\def\nn{\nonumber}

\def\Qj{Q_j^-}
\def\Qk{Q_k^+}
\def\Pj{P_j^-}
\def\Pk{P_k^+}
\def\Qjk{Q_{kj}}
\def\Pjk{P_{jk}}

\begin{document}

\title{A framework for analyzing contagion in assortative banking networks}
\author{T. R. Hurd}
\affiliation{Department of Mathematics, McMaster University, Canada}
\author{J. P. Gleeson}
\affiliation{MACSI, Department of Mathematics \& Statistics, University of Limerick, Ireland}
\author{S. Melnik}
\affiliation{MACSI, Department of Mathematics \& Statistics, University of Limerick, Ireland}

\begin{abstract}
We introduce a probabilistic framework that represents stylized banking networks with the aim of predicting the size of contagion events. Most previous work on random financial networks assumes independent connections between banks, whereas our framework explicitly allows for (dis)assortative edge probabilities (e.g., a tendency for small banks to link to large banks). We analyze default cascades triggered by shocking the network and find that the cascade can be understood as an explicit iterated mapping on a set of edge probabilities that converges to a fixed point. We derive a cascade condition that characterizes whether or not an infinitesimal shock to the network can grow to a finite size cascade, in analogy to the basic reproduction number $R_0$ in epidemic modelling. The cascade condition provides an easily computed measure of the systemic risk inherent in a given banking network topology. Using the percolation theory for random networks we also derive an analytic formula for the frequency of global cascades. Although the analytical methods are derived for infinite networks, we demonstrate using Monte Carlo simulations the applicability of the results to finite-sized networks. We show that {\it edge-assortativity}, the propensity of nodes to connect to similar nodes, can have a strong effect on the level of systemic risk as measured by the cascade condition. However, the effect of assortativity on systemic risk is subtle, and we propose a simple graph theoretic quantity, which we call the {\it graph-assortativity coefficient}, that can be used to assess systemic risk.
\end{abstract}
\maketitle

\section{Introduction}
\label{sec:1}
The study of contagion in financial systems is topical in light of the recent global credit crisis and the resultant damage inflicted on financial institutions. {\it Contagion}\cite{Hurd2016} refers to the spread of dangerous shocks through a system of financial institutions, with each successive shock causing increasing pressure on the remaining components of the system. The term {\it systemic risk} refers to the contagion-induced threat to the financial system as a whole, due to the failure of one (or more) of its component institutions.

Over time, the nature of such contagious shocks has been a topic of active discussion, with a growing list of channels such as funding illiquidity, asset fire sales and collateral shocks, that extend beyond the standard default or insolvency channel. For clarity in this paper, we adhere to the main body of systemic risk modelling, and focus only on the default channel.

It is widely held that financial systems (see Ref.~\onlinecite{Upper11} and references therein), defined for example as the collection of banks and financial institutions in a developed country, can be modelled as a random network of {\it nodes} or {\it vertices} with stylized balance sheets, connected by directed links or edges that represent exposures or {\it interbank loans}, each edge with a positive weight that represents the size of the exposure. If ever a node becomes {\it insolvent} and ceases to operate as a bank, it will create balance sheet shocks to other nodes, creating the potential of chains of insolvency that we will call {\it default cascades}. Financial networks are difficult to observe because interbank data is often not publicly available, but studies have indicated that they share characteristics of other types of technological and social networks, such as the World Wide Web and Facebook. For example, the degree distributions of financial networks are thought to be {\it fat-tailed} since a significant number of banks are very highly connected. A less studied feature observed in financial networks (and as it happens, also the World Wide Web) is that they have  high negative  {\it assortativity} \cite{Soramakietal07}. This refers to the property that any bank's counterparties (i.e., their graph neighbours) have a tendency to be banks of an opposite character. For example, it is observed that small banks tend to link preferentially to large banks rather than other small banks. Commonly, social networks are observed to have positive rather than negative assortativity. Structural characteristics such as degree distribution and assortativity are felt to be highly relevant to the propagation of contagion in networks but the nature of such relationships is far from clear~\cite{MayArin10}.

Our aim here is to develop a mathematical framework that will be able to determine the systemic susceptibility in a rich class of infinite random network models with enough flexibility to include the most important structural characteristics of real financial networks, in particular with general degree distributions and a prescribed edge-assortativity. Our starting point will be the Gai-Kapadia (GK) cascade model~\cite{GaiKapa10} and the analytical methods developed there and in Ref.~\onlinecite{GleHurMelHac12} for that model. The basic assumptions introduced in the GK model are:
\begin{enumerate}
  \item The network is a large (actually infinite) random directed graph with a prescribed degree distribution;
  \item Each node (bank) is labeled with a stylized banking balance sheet that identifies its external assets and liabilities, its internal (i.e., total interbank) assets and liabilities, and $\gamma$, its net worth or equity (i.e., its total assets minus its total liabilities). Initially, the system is in equilibrium, meaning each node has positive net worth $\gamma>0$;
  \item Each directed edge is labeled with a deterministic weight that represents the positive exposure of one bank to another. These weights depend deterministically on the in-degree of the edge, and are consistent with the interbank assets and liabilities at each node;
  \item A random shock is applied to the balance sheets in the system that triggers the default or insolvency of a fixed fraction of nodes;
  \item The residual value of an interbank exposure available to creditors of a defaulted bank is zero, and thus the shock has the potential to trigger a cascade of further bank defaults.
\end{enumerate}
The principle of {\it limited liability} for banks means that shareholders are never asked to cover a negative net worth of an insolvent firm. Instead, the insolvent firm is assumed to {\it default}. This means it ceases to operate as a going concern, shareholders are wiped out, and its creditors divide the residual value. Since this residual value is always less than the nominal liabilities, creditor banks thus receive a shock to their balance sheets, which creates the potential for a default cascade. The GK model makes a very simple {\it zero recovery} assumption that residual values of defaulted banks will be zero, and thus every time a bank defaults a maximal possible shock will be transmitted to its creditors.

Our paper makes the following contributions towards developing a mathematical theory of systemic risk.
\begin{enumerate}
  \item We generalize the GK model in an important respect, namely that the edge degree distribution $Q$ is arbitrary, allowing for any desired amount of assortativity in the network.
  \item We present a simple algorithm for constructing general assortative random directed graphs of the configuration class.
  \item We provide formulas for the expected cascade size, the frequency of global cascades, and the spectral {\it cascade condition}.
  \item We introduce the concept of {\it graph assortativity} for directed graphs that can be used to assess systemic risk.
\end{enumerate}

The remainder of this paper is structured as follows. In Sec.~\ref{sec:GK}, we introduce our model. In Sec.~\ref{sec:analyt_results}, we present our analytical results, including the calculation of the expected cascade size, the cascade condition, and a formula for the frequency of large scale cascades. In Sec.~\ref{sec:num}, we compare numerical results of Monte Carlo simulations with the analytical predictions of Sec.~\ref{sec:analyt_results} for several examples of networks generated using our model. Section~\ref{sec:conclusion} concludes.

\section{The Banking Network Model} \label{sec:GK}
In this section we specify the two constituent parts of our interbank model: network structure and dynamics. The structure or {\it skeleton} of the network is modelled as a random directed graph. The dynamics is determined by the bank balance sheets and the rules for the propagation of defaults through the interbank network.

\subsection{The Assortative Skeleton Network} \label{sec:skel}
The first step in building a financial network is to build the skeleton random directed graph where nodes represent banks and edges represent interbank loans. Our construction is an extension of the well-known configuration graph model~\cite{Bollobas01}, and to describe it we introduce the following definitions and notation:
\begin{enumerate}
  \item A node $v$ has type $(j,k)$ means its {\it in-degree}, the number of in-pointing edges, is $j$ and its {\it out-degree} is $k$.
  \item An edge $\ell$ is said to have type $(k,j)$ with out-degree $k$ and in-degree $j$  if it is an  out-edge of a node with out-degree $k$ and an in-edge of a node with in-degree $j$.
  \item We write $\CE^+_v$ (or $\CE^-_v$) for the set of out-edges (respectively, in-edges) of a given node $v$. We write $v^+_\ell$ (or $v^-_\ell$) for the node for which $\ell$ is an out-edge (respectively, in-edge). In other words, edge $\ell$ starts from $v^+_\ell$ and ends at $v^-_\ell$.
  \item Let $P_{jk}$ be the probability of a type $(j,k)$ node. This distribution has marginals $P^{+}_k:=\sum_j P_{jk}$ and $P^{-}_j:=\sum_kP_{jk}$, and mean in- and out-degree $z=\sum_jjP^{-}_j=\sum_k kP^{+}_k$.
  \item Let $Q_{kj}$ be the probability of a type $(k,j)$ edge. This distribution has marginals $Q^{+}_k:=\sum_j Q_{kj}$ and $Q^{-}_j:=\sum_kQ_{kj}$.
\end{enumerate}

\begin{figure}[h]
\includegraphics[width=0.77\columnwidth]{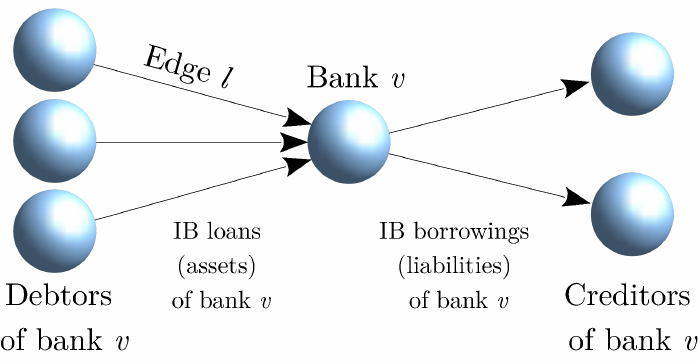}
\caption{The network neighborhood of a bank $v$ which has type $(j=3,k=2)$, since it has $3$ debtors and $2$ creditors in the interbank network. Edge $\ell$ has type $(k=1,j=3)$, since it is an out-edge of a node with out-degree $1$ and an in-edge of a node with in-degree $3$.}
\label{figS1}
\end{figure}
Figure~\ref{figS1} illustrates the neighborhood of a type $(j=3,k=2)$ node. Arrows point from debtor to creditor banks, so that default contagion propagates along the edge directions.

To define an ensemble of directed configuration graphs with $N$ nodes and joint distributions of node types $P$ and edge types $Q$ the following {\it consistency conditions} should hold for each $j$ and $k$
\begin{align}\begin{split}
\text{\textbullet} & \; NP_{jk}~\in ~\mathbb{Z},~NzQ_{kj}~\in ~\mathbb{Z}, \\
\text{\textbullet} & \; Q^{+}_k=kP^{+}_k/z, \quad Q^{-}_j=jP^{-}_k/z. \label{consistency}
\end{split}\end{align}
Here, the first condition states that there must be an integer number of nodes and edges, while the second condition ensures that the number of edges of different types corresponds exactly to the degrees of nodes. Under these conditions, we use the following algorithm to construct a directed edge-assortative graph from our ensemble:
\begin{enumerate}
  \item Make a list of $N$ nodes of which exactly $NP_{jk}$ are of type $(j,k)$ and a list of $zN$ edges of which exactly $NzQ_{kj}$ have type $(k,j)$. We refer to the unpaired in (out) arrows of each node and edge as {\it $j$-stubs} (or {\it $k$-stubs}).
  \item While there are unmatched stubs
  \begin{itemize}
    \item Pick an unmatched edge at random. Let its type be $(k,j)$.
    \item Match its $j$-stub to a random unpaired $j$-stub of a node, chosen uniformly at random from unmatched $j$-stubs.
    \item Match its $k$-stub to a random unpaired $k$-stub of a node, chosen uniformly at random from unmatched $k$-stubs.
  \end{itemize}
\end{enumerate}

\begin{figure}[h]
\includegraphics[width=0.7\columnwidth]{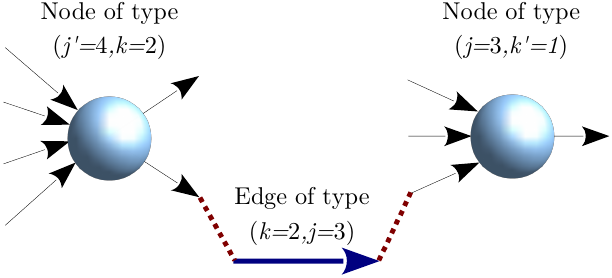}
\caption{Stub-matching during the network construction process. The $k$-stub of an unmatched edge of type $(k=2,j=3)$ is matched to an unmatched $k$-stub of one node, while the $j$-stub of the edge is matched to an unmatched $j$-stub of another node.}
\label{Construction}
\end{figure}

Two recent papers, Refs.~\onlinecite{DeprWuth15t} and \onlinecite{Hurd16b}, have explored the class of Assortative Configuration Graphs, and proposed more complex simulation algorithms that do not rely on the rationality of $P,Q$ in \eqref{consistency}. The algorithm we propose here is easy to understand, and adequate for our purposes.

We illustrate the stub-matching process in Fig.~\ref{Construction}. It is important to recognize that this graph construction may lead for finite $N$ to self-edges as well as multiple edges between node pairs. Such anomalies do not seriously affect finance interpretations and occur with vanishing probability as $N \to \infty$. The property of configuration graphs in the $N\to\infty$ limit, called the {\it locally tree-like (LT) property}, is that cycles of any fixed finite length occur only with zero probability: This has been recently proven in Ref.~\onlinecite{Hurd16b}.

The special case $Q_{kj}=kjP^{-}_jP^{+}_k/z^2=Q^{-}_jQ^{+}_k$ corresponds to edge uncorrelated directed graphs where in and out degrees of an edge are independent from each other. Such graphs can be constructed using a simpler algorithm: one lists $j$ and $k$-stubs of all nodes, and then $j$-stubs are matched to $k$-stubs uniformly at random. We are interested in the general assortative case described above because real financial networks appear to have negative edge-assortativity, in that high degree banks attach preferentially to low degree banks~\cite{Soramakietal07}.

A natural measure of edge-assortativity by degree is the {\it edge-assortativity coefficient} $r_Q\in[-1,1]$ given by
\be\label{edgeassortcoef}
r_Q=\frac{\sum_{jk}jk[Q_{kj}-Q^-_jQ^+_k]}{\sqrt{\left(\sum_{j}j^2Q^-_j-(\sum_{j}jQ^-_j)^2\right)\left(\sum_{k}k^2Q^+_k-(\sum_{k}kQ^+_k)^2\right)}}\ .
\ee
This is of course the Pearson correlation for $Q_{kj}$ viewed as a bivariate probability distribution. We will soon find some evidence that systemic risk of a network may be more strongly related to a combination of edge- and node-assortativity (arising from the dependence between in- and out- degrees of nodes). We therefore also define a measure we call the {\it graph-assortativity coefficient} $r\in[-1,1]$ given by
\be\label{graphassortcoef}
  r=\frac{\sum_{jj'}jj'[B_{jj'}-B^-_jB^+_{j'}]}{\sqrt{\left(\sum_{j}j^2B^-_j-(\sum_{j}jB^-_j)^2\right)\left(\sum_{j'}{j'}^2B^+_{j'}-(\sum_{j'}j'B^+_{j'})^2\right)}}\ ,
\ee
where
\beqq
B_{jj'}
&=&\sum_k\frac{P_{jk}}{P^+_k}Q_{j'k}
\eeqq
is the joint distribution of the in-degree of pairs of nodes connected by an edge and $B^-_j=\sum_{j'}B_{jj'}, B^+_{j'}=\sum_j B_{jj'}$ are the marginals.

\subsection{Contagion Dynamics}
\begin{figure}[h]
\includegraphics[width=0.77\columnwidth]{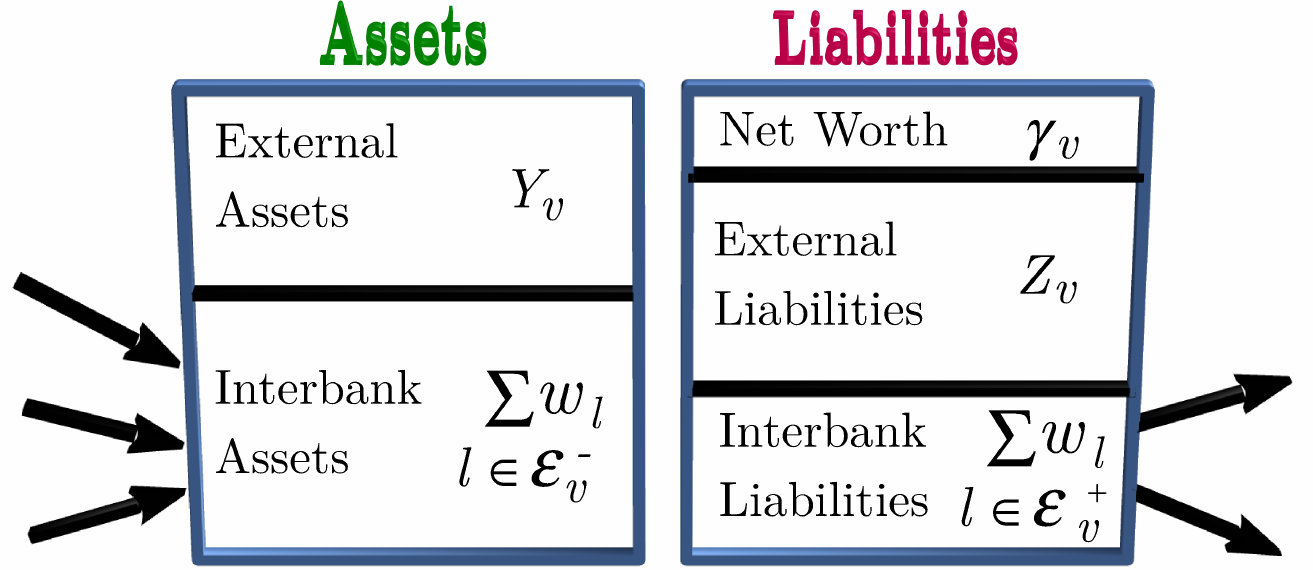}
\caption{Balance sheet of a bank $v$ with 3 debtors and 2 creditors. The {\it net worth} (or capital buffer) $\gamma_v$ of the bank is its assets less its liabilities.}
\label{figS2}
\end{figure}

To build a financial network with full accounting information, consistent with a given skeleton graph, one specifies the external assets $Y_v$ and external liabilities $D_v$ for each node $v$, and for each edge $\ell$ of the network, an exposure size or weight $w_\ell$. Then the interbank assets are $Z_v=\sum_{\ell\in\CE^-_v} w_\ell$ and interbank assets are $X_v=\sum_{\ell\in\CE^+_v} w_\ell$.  The {\it net worth} or {\it equity} of a node $v$ is defined to be its total assets minus total liabilities:
\begin{equation}
\label{networth}
\gamma_v=Y_v+Z_v-D_v-X_v\; .
\end{equation}
In Fig.~\ref{figS2}, we show the schematic balance sheet. By limited liability, the {\it solvency condition} for a bank $v$ is $\gamma_v>0$.  We will always assume that the system is initially in an equilibrium state in which all banks are solvent. Thus $\gamma_v$ is a {\it capital buffer} that keeps the bank solvent when subjected to balance sheet shocks up to a certain size.

The cascade dynamics that we specify below do not depend on full accounting information, but only on the information about the buffers $\gamma$ and edge weights $w$. The analytical results of our paper hold for default buffers $\gamma_{jk}$ that may depend on the node type $(j,k)$, and the edge weights $w_j$ that may depend on the edge in-degree.

Insolvencies arise in a system initially in equilibrium only when a shock causes at least one node to suffer a loss larger than its buffer $\gamma_v$. For simplicity, we suppose that such an initial shock to our system causes an initial set $\CMini$ of nodes to become insolvent (for example by hitting their external assets), but leaves other banks' balance sheets unchanged. The set $\CMini$ is drawn randomly, with the fraction of type $(j,k)$ nodes that are defaulted denoted by $\rhoini_{jk}$.

Under the {\it zero recovery} assumption that an insolvent bank can pay none of its interbank credit obligations, each insolvent node $v$ triggers all its out-edges to have zero value. This triggering of edges to default is an instance of what we call an {\it edge update} step of the cascade: for any set of defaulted nodes $\CM$ we find a default edge set $\CD$ which is composed only of edges originating from nodes $\CM$.

Each such defaulted edge $\ell$ now transmits a maximal shock $w_\ell$ to the asset side of the balance sheet of its end-node $v^-_\ell$ (the creditor bank). A solvent bank becomes defaulted if the total shock received by the bank from all its defaulted debtors exceeds its buffer. Hence the insolvency condition on a $(j,k)$-type node $v$ is
$$ \gamma_{jk} \le \sum_{\ell\in\CE^-_v}{\bf 1}_{\{\ell\in\CD\}} \,w_j \;  ,$$
where $\CD$ is a set of defaulted edges, and the indicator function ${\bf 1}_A$ of a set $A$ is $1$ on the set and $0$ on its complement. We call this triggering of nodes to default a {\it node update} step of the cascade: for the default edge set $\CD$ we find a default node set $\CM'$ defined by the condition $v\in\CM'$ if and only if
\be\label{nodeupdate}
\#\{\CE^-_v\cap\CD\}\ge  M_{jk}:=\lceil \gamma_{jk}/w_j\rceil  \; ,
\ee
where $(j,k)$ is the type of node $v$. Here $\lceil x\rceil $ denotes the {\it ceiling function}, i.e., the smallest integer greater than or equal to  $x$, and so $M_{jk}$ is the threshold for the number of defaulted in-edges that will cause a type $(j,k)$ node to default.

To summarize, our banking system is specified by a skeleton random directed graph (defined by the number of nodes $N$ and the probabilities $P_{jk}$, $Q_{kj}$ for node and edge types), the accounting information (bank default buffers $\gamma_{jk}$ and interbank loan amounts $w_j$) and the initial default probabilities $\rhoini_{jk}$ for each bank type (resulting in the randomly-drawn initial shocked set $\CMini$). Given any realization of a shocked financial system so specified, the default cascade will be an alternating sequence of edge and node updates, beginning with $\CMini$.

\section{Analytical results}\label{sec:analyt_results}
\subsection{Expected cascade size}\label{sec:cascade}
In this section, we calculate the expected fraction of defaulted nodes and edges in an asymptotically large network. Given any realization of a shocked financial system as specified above, with an initial shocked set $\CMini$, the default cascade can be thought of as a sequence of updates:
\begin{displaymath}
    \xymatrix{
          & \CD_1 \ar[d]& \CD_2 \ar[d]& \CD_3 \ar[d]\\
        \CMini    \ar[ur]  & \CM_1\ar[ur]   & \CM_2\ar[ur] &\CM_3\ \cdots }
\end{displaymath}
Inductively, we have nondecreasing sequences of sets for $n\ge1$:
\begin{align}
\CD_n&:=\mbox{defaulted edges triggered by nodes in $\CM_{n-1}$},\\
\CM_n&:=\mbox{defaulted nodes triggered by edges in $\CD_{n}$}.
\end{align}

We define $\rho^{(n)}_{jk}$ as the probability that a type $(j,k)$ node is in the default set $\CM_n$, and probabilities $\sigma^{(n)}_k$ and $a^{(n)}_j$ that respectively an edge with out-degree $k$ and an edge with in-degree $j$ are in the default set $\CD_n$. To calculate these probabilities, we use a simple but powerful recursive approach for solving cascade-type dynamics on random network models~\cite{Gleeson08a, Melnik13, Melnik14}.

Consider a type $(j,k)$ node and calculate its default probability $\rho_{jk}^{(n)}$ for $n\ge 1$. The node is either initially defaulted with probability $\rhoini_{jk}$, or it is initially not defaulted with probability $1-\rhoini_{jk}$. In the latter case, it will default if it has sufficiently many defaulted in-edges. Each of its $j$ in-edges is defaulted with probability $a_j^{(n)}$. From the locally tree-like property of the skeleton in the limit $N\to\infty$, we deduce that the states of the in-edges of a node are independent from each other. Therefore, the probability of exactly $m$ out of $j$ in-edges to be is the binomial probability  $\dbinom{j}{m} (a_j^{(n)})^m(1-a_j^{(n)})^{j-m}$. These $m$ defaulted edges cause the default of the node if $m$ is at least $M_{jk}=\lceil \gamma_{jk}/w_j\rceil$ (see Eq.~\eqref{nodeupdate}). Hence, adding all probabilities together gives
\begin{align}
\label{rhon} \rho_{jk}^{(n)} = \rhoini_{jk} + (1-\rhoini_{jk})\sum_{m= M_{jk}}^j\dbinom{j}{m} (a_j^{(n)})^m(1-a_j^{(n)})^{j-m}.
\end{align}

Next, to calculate $\sigma_k^{(n+1)}$, the probability that an edge with out-degree $k$ is defaulted at step $n+1$, we take an edge with out-degree $k$ and look at its source node which (by the definition) has out-degree $k$. This is a type $(j,k)$ node with conditional probability $P_{jk}/P^+_{k}$ and if so, it is defaulted at step $n$ with probability $\rho^{(n)}_{jk}$. Hence,
\begin{align}
\label{sign} \sigma_k^{(n+1)}=\sum_{j}\rho_{jk}^{(n)}\frac{P_{jk}}{P^+_{k}},
\end{align}
where the sum is over possible in-degrees $j$ of the source node.

Similarly, the probability that an edge with in-degree $j$  is defaulted  at step $n+1$ is given by
\begin{align}
 \label{an} a_j^{(n+1)}=\sum_k \sigma_k^{(n+1)} \frac{Q_{kj}}{Q^-_j},
\end{align}
where $Q_{kj}/Q^-_j$ is the probability that the edge has out-degree $k$, given its in-degree is $j$. An edge of type $(k,j)$ is defaulted with probability $\sigma_k$ and we sum over all possible $k$.

Starting with a given fraction of initially defaulted nodes $\rhoini_{jk}$, we begin by computing the collections $\sigma_k^{(1)}, a_j^{(1)}$ using  \eqref{sign} and \eqref{an} . Thereafter, we can iterate Eqs.~\eqref{rhon}-\eqref{an} to obtain the values of $\rho^{(n)}_{jk}$, $\sigma^{(n+1)}_k$, $a^{(n+1)}_j$,  for $n\ge1$.

In the case of edge-uncorrelated directed networks when $Q_{kj}=Q^+_k Q^-_j$, the quantities $a^{(n)}_j$ no longer depend on $j$ and Eqs.~\eqref{rhon}-\eqref{an} simplify to
\begin{align}
\rho_{jk}^{(n)} &= \rhoini_{jk}+(1-\rhoini_{jk})\sum_{m= M_{jk}}^j\dbinom{j}{m} (a^{(n)})^m(1-a^{(n)})^{j-m}\ ,\\
\label{scalarFP} a^{(n+1)}&=\sum_{j,k}\frac{k}{z}P_{jk}\rho_{jk}^{(n)}\ .
\end{align}

\subsection{The Cascade Condition}\label{sec:casccond}
We can derive a {\it cascade condition} which implies that a generic infinitesimally small fraction $\rhoini_{jk}$ of defaulted nodes will result in a cascade of finite size. Writing Eqs.~\eqref{rhon}-\eqref{an} in vector form as
\be
\label{fix} \bar a^{(n+1)}=\{G_j(\bar a^{(n)})\}\ ,
\ee
where $\bar a^{(n)} = \{a_j^{(n)}\}$, an infinitesimally small seed may only grow if  the Jacobian matrix $D_{jj'}=\partial G_j/\partial a_{j'}|_{\bf 0}$ has an expanding direction, i.e., at least one eigenvalue with magnitude bigger than 1. In Sec.~\ref{sec:num}, we shall see that the cascade condition is indeed a strong measure of systemic risk in simulated networks.

The derivatives $D_{jj'}$ are easy to calculate. From Eq.~\eqref{rhon}
\begin{align}
\nn \lb.\frac{\partial\rho^{(n)}_{jk}}{\partial a^{(n)}_j}\rb|_{a^{(n)}_j=0} & = \lb.\sum_{m= M_{jk}}^j \binom{j}{m} (a^{(n)}_j)^{m-1}(1-a^{(n)}_j)^{j-m-1}(m-a^{(n)}_j j)\rb|_{a^{(n)}_j=0} \\
\nn &= \lb.\binom{j}{M_{jk}} (1-a^{(n)}_j)^{j-M_{jk}} (a^{(n)}_j)^{M_{jk}-1} M_{jk} \rb|_{a^{(n)}_j=0}  \\
\label{aderiv} &= j {\bf 1}_{\{\gamma_{jk}\le w_j\}}.
\end{align}
Combining Eqs.~\eqref{sign} and~\eqref{an}, and substituting Eq.~\eqref{aderiv}, the linearization of $G_j(\bar a^{(n)})$ around zero is
\begin{align}
\nn G_j(\bar a^{(n)}) & = \sum_k \frac{Q_{kj}}{Q^-_j P^+_k} \sum_{j'} P_{j'k} \rho^{(n)}_{j'k}  \\
& \approx \sum_k \frac{Q_{kj}}{Q^-_j P^+_k} \sum_{j'}  P_{j'k} j' {\bf 1}_{\{\gamma_{j'k}\le w_{j'}\}} a^{(n)}_{j'},
\end{align}
which yields
\be
D_{jj'}=\lb.\frac{\partial G_j}{\partial a_{j'}}\rb|_{\bf 0}=\sum_k\frac{j'Q_{kj}P_{j'k}{\bf 1}_{\{\gamma_{j'k}\le w_{j'}\}}}{Q_j^-P_k^+}\; .
\ee
Finite size cascades are possible when the spectral radius (the largest eigenvalue in absolute value) of matrix $\{D_{jj'}\}$  exceeds one:
\begin{align}
\label{cascadecond}
||D||>1.
\end{align}

In the case of uncorrelated edge degrees (i.e., $Q_{kj}=Q^+_kQ^-_j$), $a_j$ no longer depends on $j$ and the cascade condition is simply
\begin{align}
\sum_{j,k}\frac{jk}{z} P_{jk}{\bf 1}_{\{\gamma_{jk}\le w_j\}} > 1\ ,
\end{align}
a result that has been derived previously in a rather different fashion\cite{GaiKapa10,AminContMinc11}. This formula extends the percolation theory approach from undirected networks~\cite{Watts02} to the case of directed nonassortative networks. We will see in the next section that the percolation approach to the cascade condition also extends to our directed assortative networks.

We can understand the cascade condition more clearly by introducing the notion of {\it vulnerable node}, that is any node that defaults if any one of its debtors (in-neighbours) defaults. In our specifications, a $(j,k)$ node is thus vulnerable if and only if its capital buffer is less or equal to the weight of its in-links, i.e., $\gamma_{jk}\le w_{j}$. The matrix element $D_{jj'}$ has a simple explanation that gives more intuition about the nature of the cascade condition: it is the expected number of edges with in-degree $j$ that emanate from a vulnerable node reached by following an edge with in-degree $j'$.

\subsection{Frequency of global cascades and the giant vulnerable cluster}\label{Frequency}
The cascade condition that tells us that global cascades are possible turns out to be equivalent to the existence of a giant vulnerable cluster in the interbank network. When the cascade condition is satisfied, the default of a single bank will result in a global cascade if the bank belongs to the so-called in-component of the giant vulnerable cluster. Hence, the frequency of global cascades is bounded from below (and as it turns out well approximated by) by the fractional size of the in-component (see Chapter 13.11 of Ref.~\onlinecite{Newman10}).
\begin{figure}[!h]
\centering
\includegraphics[width=0.65\columnwidth]{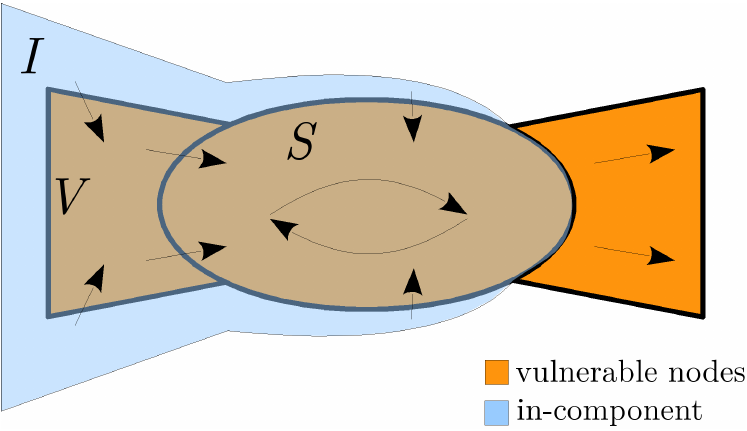}
\caption{Schematic structure of the network with arrows representing the propagation of default. The default of any bank in the in-component $I$ will trigger the default of all nodes in the strongly connected vulnerable cluster $S$, as well as the vulnerable nodes in the out-component of $S$.}
\label{fig_freq_sketch}
\end{figure}

Let us define the following (see Fig.~\ref{fig_freq_sketch}):
\begin{itemize}
  \item $\CV$ is the set of vulnerable nodes;
  \item $\CS \subset \CV$ is the giant strongly connected set of vulnerable nodes (called the {\it giant vulnerable cluster});
  \item $\CI \supset S$ is the {\it in-component} of the giant vulnerable cluster: the set of (possibly not vulnerable) nodes that are connected to $\CS$  by a directed path through vulnerable nodes;
  \item $\Gamma_{jk}={\bf 1}_{\{\gamma_{jk}\le w_j\}}$ is the indicator function that gives 1 if type $(j,k)$ nodes are vulnerable and 0 otherwise.
\end{itemize}
The default of any node in the in-component $\CI$ will cause the default of the entire strongly connected component $\CS$. We consider $\bar b=\{b_k\}$ where $b_k$ is the probability that a node with  $k$ out-neighbours is not in the in-component $\CI$. Note that $v\in \CI^c$ (i.e., the complement of $\CI$) is equivalent to  the condition that all the downstream nodes are in the set $\CV^c\cup(\CV\cap\CI^c)$, i.e., the out-neighbours of $v$ are either not vulnerable or they are vulnerable, but not in the in-component of $S$.
Thus, $b_k =(c_k)^k,$ where $c_k$ is the probability that an out-neighbor of an out-degree-$k$ node is in the set $\CV^c\cup(\CV\cap\CI^c)$.

To calculate $c_k$, we note that an out-neighbor of a type $(j,k)$ node is a $(j',k')$-type node with probability ${P_{j'k'}Q_{j'k}}/{P^-_{j'}Q^+_k}$. The probability that a $(j',k')$-type node is not vulnerable is $1-\Gamma_{j'k'}$. The probability that a $(j',k')$-type node is vulnerable, but does not belong to the in-component is $\Gamma_{j'k'}b_{k'}=\Gamma_{j'k'}(c_{k'})^{k'}$. Thus, combining all probabilities together and summing over the possible types of nodes we get
\be
\label{eq:c_k} c_k=\sum_{j',k'} \left(\Gamma_{j'k'}(c_{k'})^{k'}+(1-\Gamma_{j'k'})\right)\frac{P_{j'k'}Q_{j'k}}{P^-_{j'}Q^+_k}.
\ee
Hence, $\bar c=\{c_k\}$ can be found as a fixed point of Eq.~\eqref{eq:c_k}, which we re-write in vector form as $\bar c=\{h_k(\bar c)\}$. Note that the equation $\bar c= \{h_k(\bar c)\}$  has a trivial fixed point $\bar e=(1,1,\dots)$ that corresponds to the set $\CI$ being empty. We now verify that the cascade condition $\|D\|>1$ is equivalent to the condition that $\bar e$ is an unstable fixed point, in which case there will be a nontrivial fixed point $0\le \bar c_\infty < \bar e$. A sufficient (and almost necessary) condition for $\bar e$ to be an unstable fixed point is that $\|\tilde D\|>1$ where the derivative $\tilde D_{kk'}=(\partial h_k/\partial c_{k'})|_{\bar c=\bar e}$ is given by
\be \tilde D_{kk'}=\sum_{j'}\frac{k'Q_{j'k}P_{j'k'}\Gamma_{j'k'}}{Q^+_k P^-_{j'}}\ee
One can verify directly that
\[ \tilde D=\left(\Lambda BA\Lambda^{-1}\right)^T, \quad D=AB
\]
for matrices
\[ A_{jk}=\frac{Q_{kj}}{Q^-_j},\ B_{j'k}=\frac{j' P_{j'k}\Gamma_{j'k}}{P^+_k},\ \Lambda_{kk'}=\delta_{kk'}kP^+_k
\]
and from this it follows that the spectra, and hence the spectral radii of $\tilde D$ and $D$ are equal. Hence $\|D\|>1$ if and only if $\|\tilde D\|>1$.

As long as the cascade condition is satisfied, the cascade frequency $f$ is  approximately the lower bound given by the fractional size of the in-component $\CI$:
\be\label{frequency}
f \gtrsim \sum_k(1-(c_k)^k)P^+_k\ .
\ee

\section{Numerical Results}\label{sec:num}
In this section, we consider two examples of stylized interbank networks and show that the analytical results obtained above match well to the Monte Carlo simulations when $N$, the number of nodes in the network, is sufficiently large. Unless specified otherwise, we adopt the choice of parameters made for the model of Ref.~\onlinecite{GaiKapa10}:
$$\gamma_{jk}=\gamma:=0.035;\quad w_j=\frac1{5j}.$$

\subsection{A Simple Random Network Model}
We consider networks constructed with nodes of types $(3,3), (3,12), (12,3), (12,12)$ and edges of the same types. For parameters $a\in[0, 0.5]$ and  $b\in[0,0.2]$ the following $P$ and $Q$ matrices are consistent and specify a network with an average node degree $z=7.5$:
\begin{align}
\nn
\begin{pmatrix}
  P_{3,3}    &  P_{3,12}  \\
  P_{12,3}    &  P_{12,12}
\end{pmatrix}&=
\begin{pmatrix}
 0.5-a     &a    \\
   a   &  0.5-a
\end{pmatrix},\quad \\
\label{PQ}\\
\nn
\begin{pmatrix}
  Q_{3,3}    &  Q_{3,12}  \\
  Q_{12,3}    &  Q_{12,12}
\end{pmatrix}&=
\begin{pmatrix}
 0.2-b     &b    \\
   b   &  0.8-b
\end{pmatrix}\; .
\end{align}

We first fix the value of $a$ to be $0.5$, which means that the in- and out-degrees of all nodes are negatively correlated: nodes with in-degree 3 have out-degree 12, and vice versa. We examine three different values of the parameter $b$: the independent connections case $b=0.16$, the near maximally positive assortative case $b=0.01$ and the near  maximally negative assortative case $b=0.19$.  Note that the independent edge condition has been assumed in all previous work on such problems. We also note that with $b=0$, edges have maximally positive assortativity and link nodes of out-degree 3 to nodes of in-degree 3 only, and nodes of out-degree 12 to nodes of in-degree 12 only. In this case, the network consists of two disconnected components.

We vary the net worth $\gamma$ over the range $0$ to $0.1$, while the initial shock distribution is taken to be $\rhoini_{jk}=1/N$ for all types $(j,k)$, corresponding to the shocking of a single randomly-chosen bank.
\begin{figure}[!h]
\includegraphics[width=0.97\columnwidth]{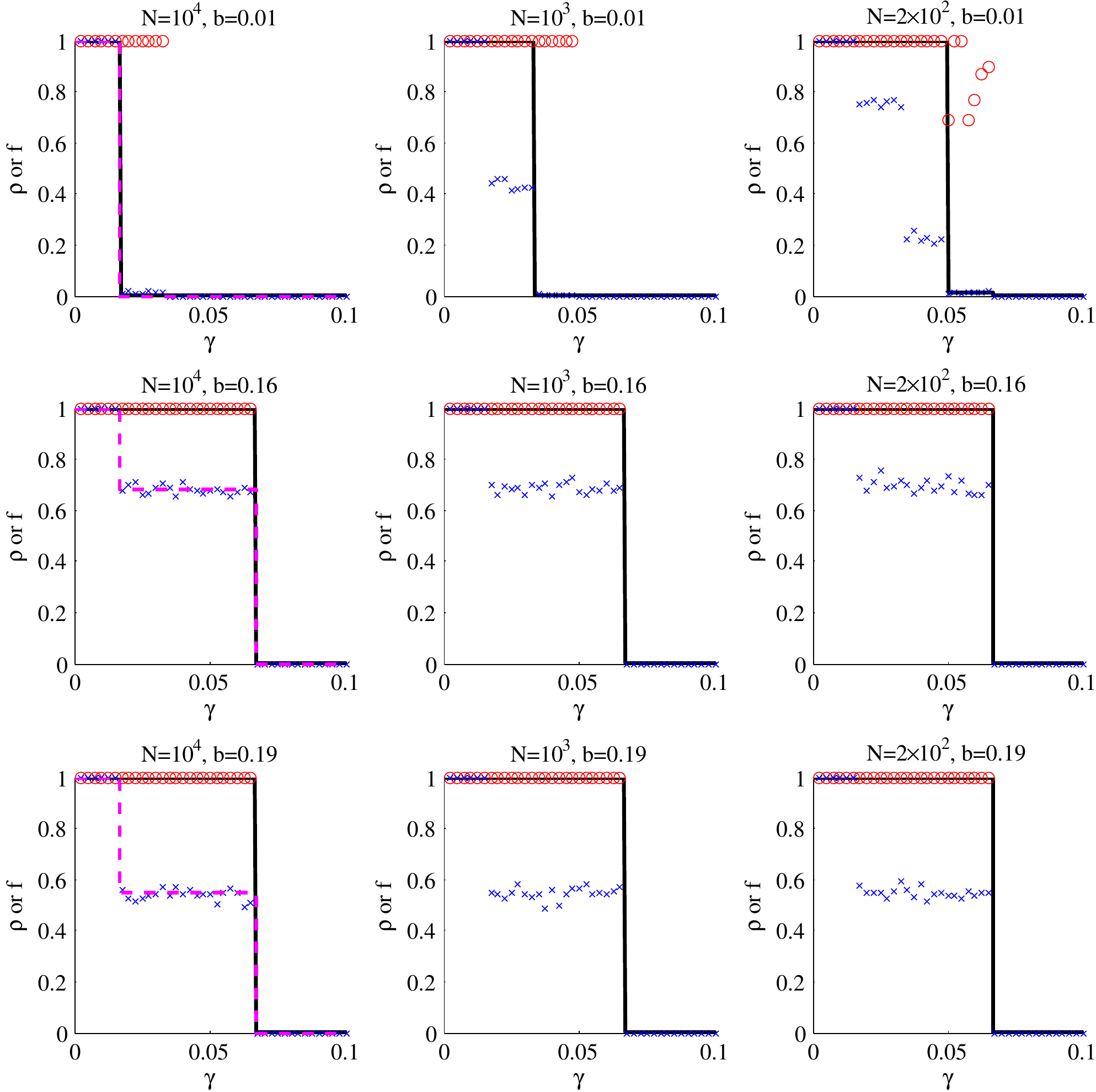}
\put(-240,230){\text{\bf(a)}}
\put(-240,150){\text{\bf(b)}}
\put(-240,70){\text{\bf(c)}}
\caption{Numerical simulation results (symbols) and theoretical results (curves) for the random network model of Eq.~\eqref{PQ}, on networks of $N$ nodes with parameter $a=0.5$, as functions of the net worth $\gamma$. The average size and frequency of global cascades in simulations are shown by red circles and blue crosses, respectively. Theoretical results for the expected cascade size (black solid curve) are from Sec.~\ref{sec:cascade}; those for the frequency of cascades (dashed magenta curve) are from Sec.~\ref{Frequency}. Each column shows results for a different network size $N$, and the parameter $b$ takes a different value on each row of the figure. Since the (dashed magenta) frequency curves are independent of $N$ they are only shown in the first column.}
\label{figR1}
\end{figure}

Figure~\ref{figR1} compares theory curves for cascade size (found by iterating Eqs.~\eqref{sign}--\eqref{rhon} to convergence) as well as the cascade frequency given by Eq.~\eqref{frequency} with results from numerical simulations on random networks with $N=10^4, 10^3$ and $200$ nodes. The node correlation parameter is fixed at $a=0.5$, while the edge correlation parameter takes the values $b\in\{0.01, 0.16, 0.19\}$. Results are plotted as functions of the net worth parameter $\gamma$. In each case, 500 realizations are used to find the extent of global cascades (a global cascade is defined, similarly to Refs.~\onlinecite{GaiKapa10,GleHurMelHac12}, as one in which more than 5\% of nodes default), and the frequency with which such global cascades occur. As expected, the analytical approach accurately predicts the size of the global cascades. Some discrepancies may be noted in Fig.~\ref{figR1}, where the theory does not predict some global cascades, but note that these occur with only very small frequencies.

The cascade condition (\ref{cascadecond}) predicts that the critical values of the cascade buffer parameter $\gamma$ are: $\gamma_c=0.017$ for the parameters of Fig.~\ref{figR1}(a), and $\gamma_c=0.067$ for the case of Fig.~\ref{figR1}(b). These values match very accurately to the locations of the dramatic transitions in the theory curve (and in the expected size of cascades in numerical simulation): for $\gamma$  values in excess of $\gamma_c$ global cascades are extremely rare, while for values  less than $\gamma_c$ the entire financial system is likely to fail following a single bank's default.  These result indicate the potential usefulness of the cascade condition as a measure of systemic risk.

\begin{figure}[htb]
\centering
\includegraphics[width=0.9\columnwidth]{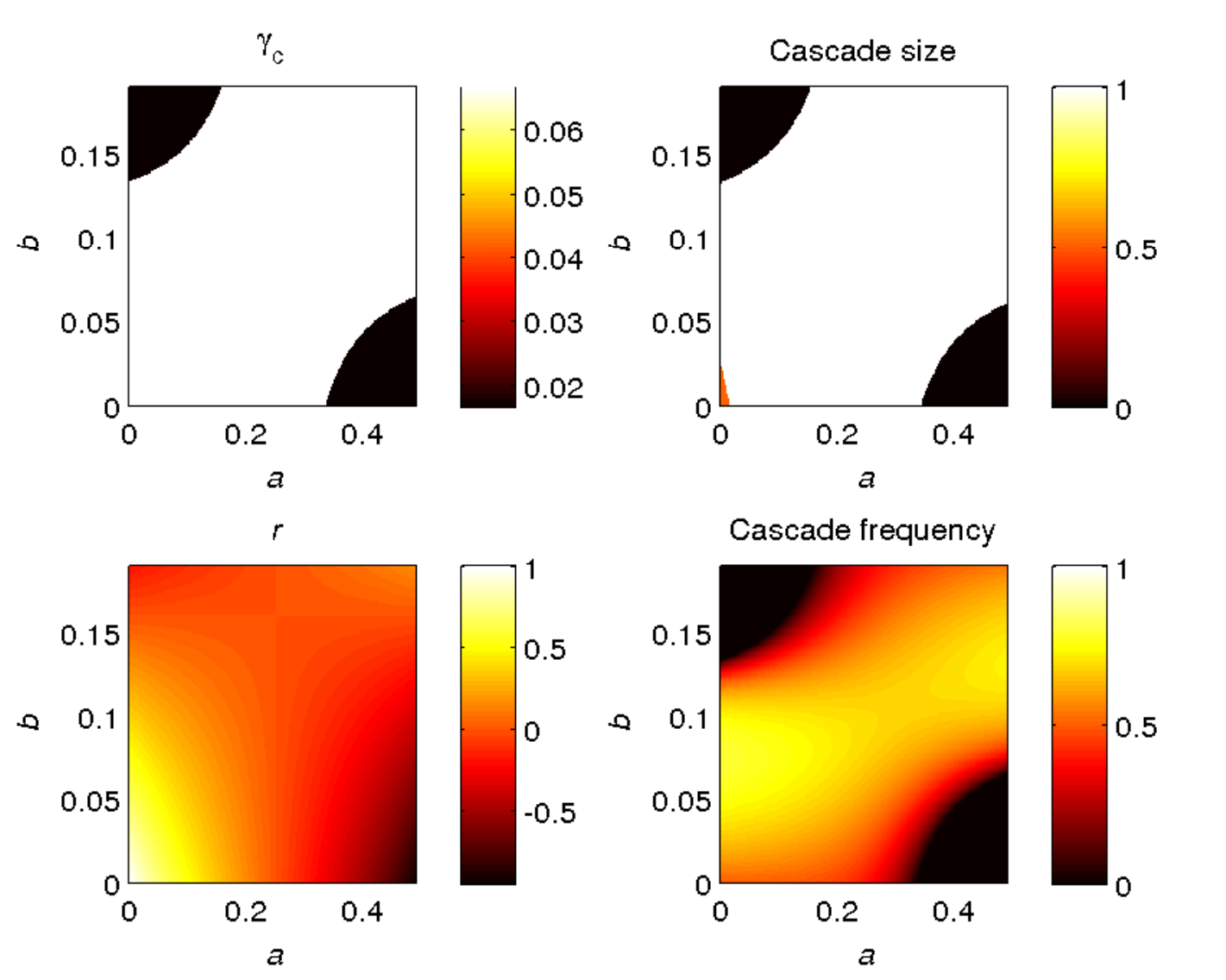}
\caption{Figure showing how various theoretical quantities of the network defined by Eqs.~\eqref{PQ} depend on the parameters $(a,b)$. Top left: critical $\gamma$ value. Top right: Expected size of cascades (from Sec.~\ref{sec:casccond}) when $\gamma=0.05$ and $\rhoini_{j k}=10^{-4}$. Bottom left: the graph assortativity parameter $r$. Bottom right: frequency of cascades (from Sec.~\ref{Frequency}) when $\gamma=0.05$.}
\label{fig_draw_SIFIN_fig2}
\end{figure}
\begin{figure}[htb]
\centering
\includegraphics[width=0.97\columnwidth]{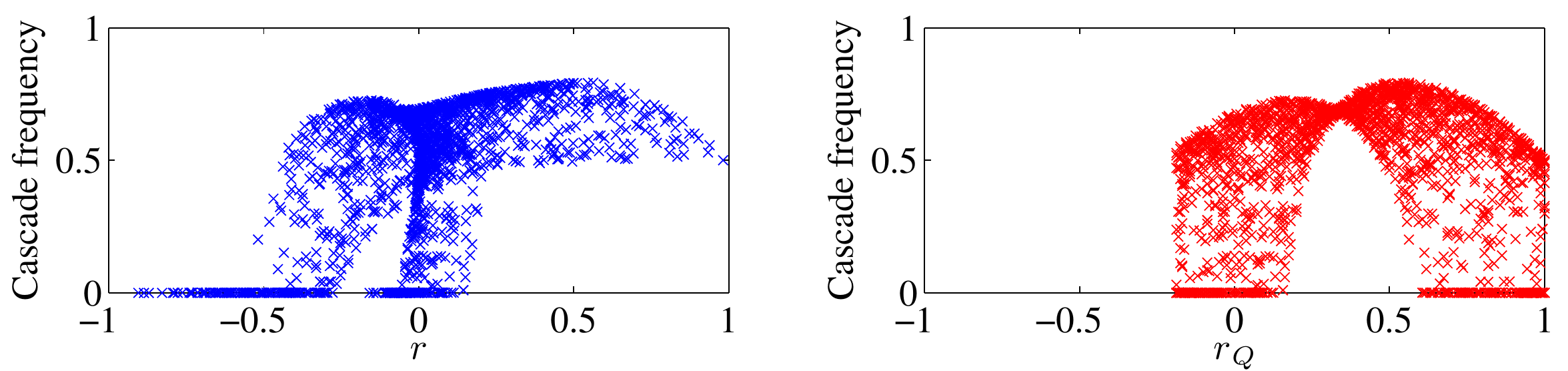}
\caption{Scatter plots showing the correlation between the cascade frequency and $r$ (left panel) and $r_Q$ (right panel) for the example shown in Fig.~\ref{fig_draw_SIFIN_fig2}. 
Each point represents a pair of (a,b) values (taken uniformly at random from the range shown in Fig.~\ref{fig_draw_SIFIN_fig2}), for which we calculate the cascade frequency and $r$ (left panel), or the cascade frequency and $r_Q$ (right panel). Note that $r$ is a better predictor of cascade frequency than $r_Q$ in this example.}
\label{fig_scatter}
\end{figure}

In Fig.~\ref{fig_draw_SIFIN_fig2}, we consider the dependences on $(a,b)$ of various theoretical quantities in the infinite $N$ limit. In the top panels, the critical value of $\gamma$ and cascade size are seen to be discontinuous, and certainly not related to edge-assortativity (which is monotonic in $b$). On the other hand (see bottom panels), the frequency of cascades is continuously varying, and does appear to correlate to some extent with the graph assortativity coefficient $r$ given by Eq.~\eqref{graphassortcoef}.  We observe in the two scatter plots of Fig.~\ref{fig_scatter} that in this model $r$ is a better purely graph theoretic predictor of systemic susceptibility than $r_Q$.

\subsection{Hierarchical Banking Network}
\begin{figure}[ht]
\centering
\includegraphics[width=0.9\columnwidth]{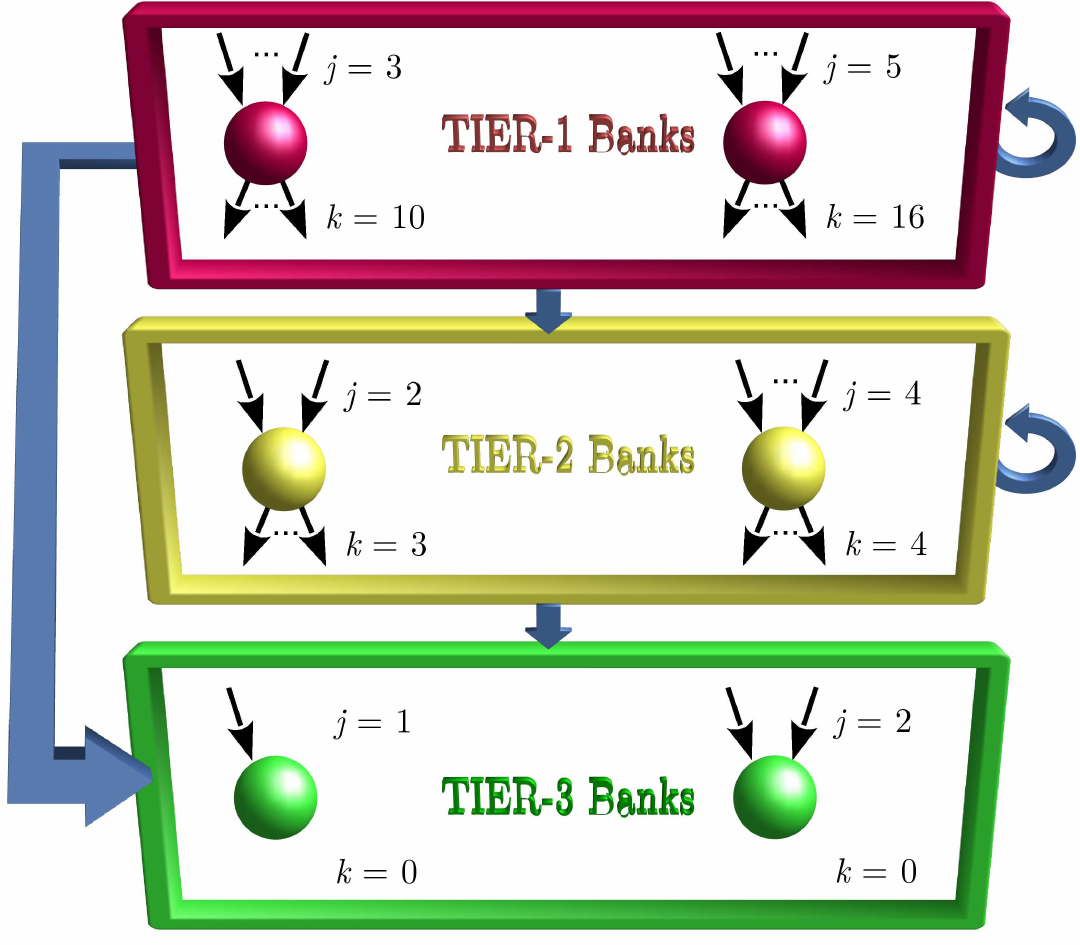}
\caption{Sketch of a directed assortative interbank network defined by Eqs.~\eqref{Pjk_mat} and~\eqref{Qjk_mat}. The network consists of 3 tiers of banks. The connections among tiers are shown by the thick arrows and represent possible paths for the spread of defaults. The default of Tier-3 banks cannot cause any further bank defaults because Tier-3 banks have 0 out-degree. Tier-2 banks can cause the default of Tier-2 and Tier-3 banks. The default of a Tier-1 bank may lead to the default of banks in any tier. Tier-1 banks consist of nodes of types (3,10) and (5,16), Tier-2 banks of nodes (2,3) and (4,4), and Tier-3 banks of nodes (1,0) and (2,0).}
\label{Network_sketch}
\end{figure}

It is known that small banks often tend to be net lenders in the interbank sector, while large banks tend to be net interbank borrowers. One also expects small banks to have few counterparties while large banks have many. In Fig.~\ref{Network_sketch}, we sketch a proposed stylized interbank network with banks divided into three tiers: small Tier-3 banks, medium Tier-2 banks and large Tier-1 banks.
We suppose that Tier-3 banks typically do not borrow from other banks, and deposit their excess funds in one or two Tier-2  or Tier-1 banks. Tier-2 banks may borrow from two or three Tier-3 banks and one or two Tier-2 banks, while they lend (deposit) to several Tier-2 or Tier-1 banks. Finally, we suppose that Tier-1 banks borrow from a handful of Tier-3 banks, several Tier-2 and Tier-1 banks. Note that one needs our assortative model to represent a tiered interbank network sketched in Fig.~\ref{Network_sketch}, as the previously developed models\cite{GaiKapa10} are unable to do so. The following $P$ and $Q$ matrices realize these characteristics in our model:
\def\vsp{\vspace{4pt}}
\begin{align}
  \begin{tabular}{c@{}c@{\enskip}c@{\enskip}c@{\enskip}c@{\enskip}c@{ }ccllc}
    &$k:$ &$0$ & $3$& $4$ & $10$ & $16$ &\vspace{10pt} \\
    &\multirow{6}{*}{$\lb.\begin{array}{c}\\ \\ \\ \\ \\ \\ \\ \end{array}\rb($} 
    & 40& 0 & 0 & 0 & 0&
    \multirow{6}{*}{$\lb) \begin{array}{c}\\ \\ \\ \\ \\ \\ \\ \end{array}\rb.$} 
    &{$j=1$} \vsp\\
    && 30 & 10 & 0 & 0 & 0&& $j=2$ \\
    $\Pjk = \displaystyle\frac{1}{100}$
    && 0 & 0 & 0 & 5 & 0&& $j=3$ \\
    && 0 & 0 & 10 & 0 & 0&& $j=4$ \vsp \\
    && 0 & 0 & 0 & 0 & 5&& $j=5$
  \end{tabular}
\label{Pjk_mat}
\end{align}
\begin{align}
  \begin{tabular}{c@{}c@{\enskip}c@{\enskip}c@{\enskip}c@{\enskip}c@{ }ccllc}
    &$k:$ &$0$ & $3$& $4$ & $10$ & $16$ &\vspace{10pt} \\
    &\multirow{6}{*}{$\lb.\begin{array}{c}\\ \\ \\ \\ \\ \\ \\ \end{array}\rb($} 
    & 0 & 9 & 12 & 11 & 16 &
    \multirow{6}{*}{$\lb) \begin{array}{c}\\ \\ \\ \\ \\ \\ \\ \end{array}\rb.$} 
    &{$j=1$} \vsp\\
    && 0 & 18 & 24 & 22 & 32 && $j=2$ \\
    $\Qjk = \displaystyle\frac{1}{240}$
    && 0 & 0 & 0 & 6 & 12 && $j=3$ \\
    && 0 & 9 & 12 & 11 & 16 && $j=4$ \vsp \\
    && 0 & 0 & 0 & 10 & 20 && $j=5$
  \end{tabular}
\label{Qjk_mat}
\end{align}
Here the column index corresponds to possible out-degrees $k \in \{0, 3, 4, 10, 16\}$ and the row index corresponds to possible in-degrees $j \in \{1, 2, 3, 4, 5\}$. For example, $P_{1,0} = 0.4$ means 40\% of nodes have in-degree 1 and out-degree 0, and $Q_{2,4} = 0.1$ means 10\% of edges start from nodes with our-degree 4 and end at nodes with in-degree 2. The Tier-1 banks are composed of types (3,10) and (5,16) nodes, Tier-2 banks of types (2,3) and (4,4) nodes, and Tier-3 banks of types (1,0) and (2,0) nodes. One can check that the row and column constraints $\Qj = j \Pj/z$, $\Qk = k \Pk/z$ are satisfied with mean degree $z = \sum_k k \Pk = \sum_j j \Pj = 2$.

It will be instructive to compare the default cascades on such hierarchical network with cascades on its edge uncorrelated version, i.e., on a network where in and out degrees of an edge are independent. Thus, in the edge uncorrelated case, $\Qjk$ factorizes as
\begin{align}
 \Qjk^{\rm unc} = \frac{j \Pj k \Pk}{z^2},
\end{align}
and using the values from Eq.~\eqref{Pjk_mat} one obtains
\begin{align}
    \begin{tabular}{c@{}c@{\enskip}c@{\enskip}c@{\enskip}c@{\enskip}c@{ }ccllc}
        &$k:$ &$0$ & $3$& $4$ & $10$ & $16$ &\vspace{10pt} \\
        &\multirow{6}{*}{$\lb.\begin{array}{c}\\ \\ \\ \\ \\ \\ \\ \end{array}\rb($} 
        & 0 & 24 & 32 & 40 & 64 &
        \multirow{6}{*}{$\lb) \begin{array}{c}\\ \\ \\ \\ \\ \\ \\ \end{array}\rb.$} 
        &{$j=1$} \vsp\\
        && 0 & 48 & 64 & 80 & 128 && $j=2$ \\
        $\Qjk^{\rm unc} = \displaystyle\frac{1}{800}$ && 0 & 9 & 12 & 15 & 24 && $j=3$ \\
        && 0 & 24 & 32 & 40 & 64 && $j=4$ \vsp \\
        && 0 & 15 & 30 & 25 & 40 && $j=5$
    \end{tabular}
\label{Qjk_mat_uncorr}
\end{align}
Observe that unlike~\eqref{Qjk_mat},~\eqref{Qjk_mat_uncorr} allows edges between all banks, irrespective of their degrees, so there is no hierarchical structure of Fig.~\ref{Network_sketch} in this case.

We consider directed networks generated according to $P$ matrix~\eqref{Pjk_mat}, and $Q$ matrix~\eqref{Qjk_mat} for edge correlated, or~\eqref{Qjk_mat_uncorr} for edge uncorrelated case. For simplicity, we assume as before that the default buffer $\gamma$ is the same for all nodes, and link weights are given by $1/(5j)$, where $j$ is the link in-degree. We consider scenarios under which a single bank becomes defaulted, thereby initiating a cascade of defaults.

\def\panelheight{0.3\columnwidth}
\begin{figure}[ht!]
\includegraphics[width=0.49\columnwidth]{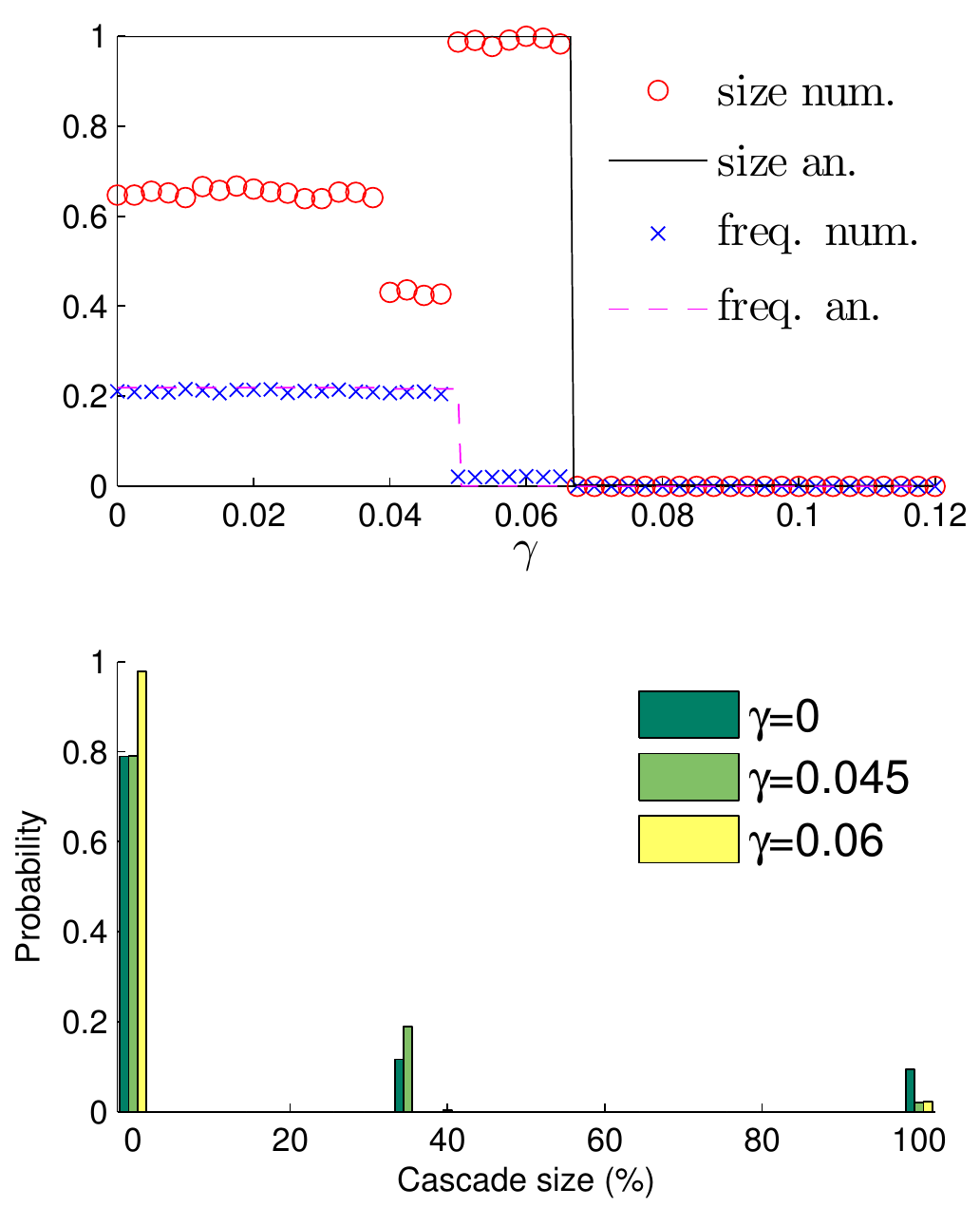}
\put(-100,135){\text{\bf(a)}}
\put(-100,60){\text{\bf(c)}}
\includegraphics[width=0.49\columnwidth]{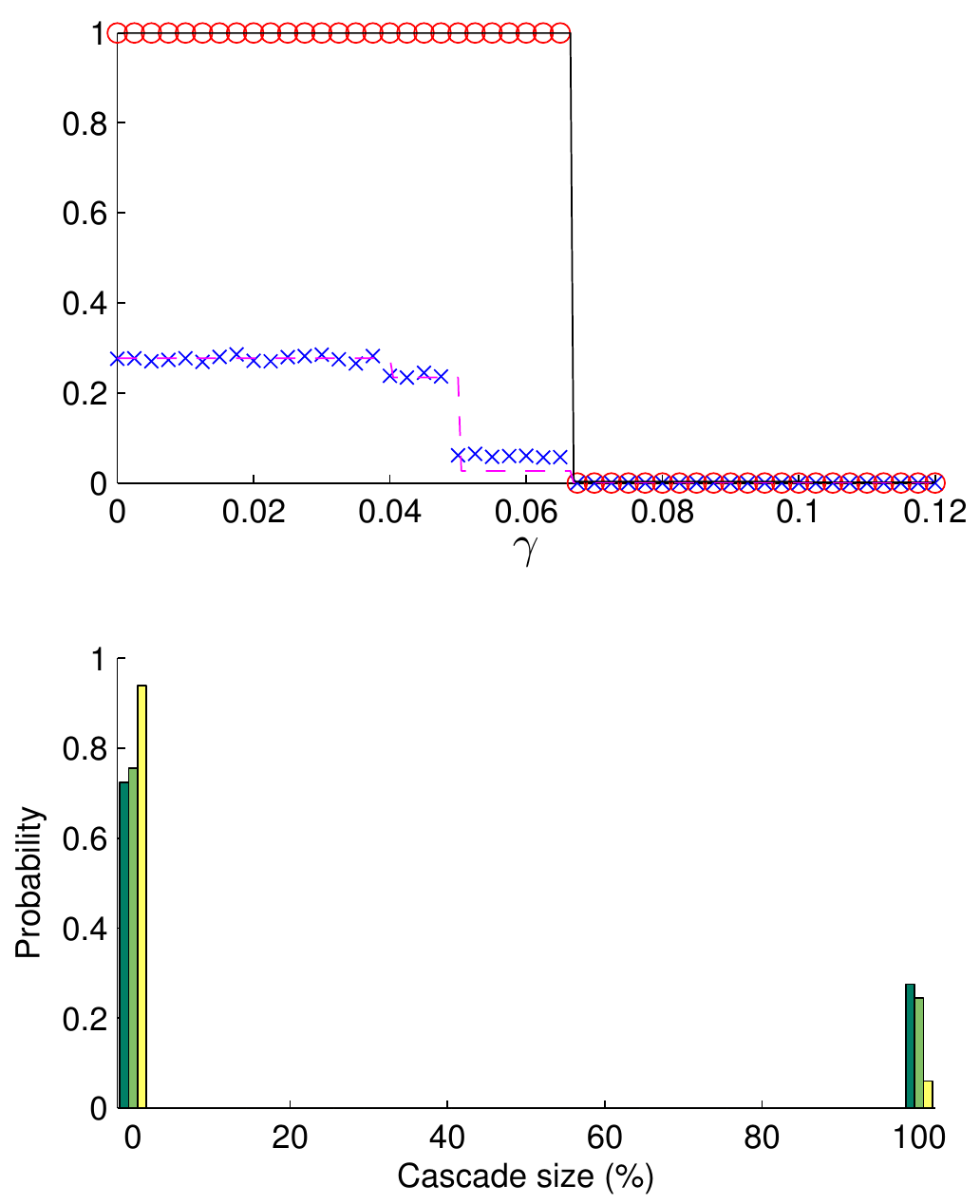}
\put(-100,135){\text{\bf(b)}}
\put(-100,60){\text{\bf(d)}}
\caption{Results for network defined by Eqs.~\eqref{Pjk_mat} and~\eqref{Qjk_mat} (left panels) and its uncorrelated version defined by Eqs.~\eqref{Pjk_mat} and~\eqref{Qjk_mat_uncorr} (right panels). Initially a single bank chosen at random from the network of $N=12000$  nodes is defaulted. To obtain analytical results we set $\rho^0_{jk} = 1/N$ (for all $j$ and $k$). Top panels show the analytical and numerical results for the expected size of global cascades and their frequency versus the default buffer $\gamma$. Bottom panels show numerically calculated distributions of cascade sizes for different values of default buffer $\gamma$. To obtain numerical results we averaged over $10^4$ realizations of random seeds, and a global cascade occurs if it occupies over 5\% of the network.}
\label{PQ45}
\end{figure}

In Fig.~\ref{PQ45} (top panels), we plot analytical and numerical results for the expected size of global cascades and their frequency versus the default buffer $\gamma$. The analytical results for the frequency and the expected cascade size were obtained using Eq.~\eqref{frequency} and Eqs.~\eqref{sign}-\eqref{rhon} respectively. In general, our theory predicts numerical results quite well. However, for some values of $\gamma$, the cascade size is not captured accurately by the theory in the correlated case shown in Fig.~\ref{PQ45}(a). This is because Eqs.~\eqref{sign}-\eqref{rhon}, as many other theoretical approaches, assume infinite network size, which is not the case here. Hence, for some parameters the theory may not perform well on finite systems \cite{Faqeeh15a, Melnik11, Gleeson12}.

In the bottom panels of Fig.~\ref{PQ45}, we show numerical distributions of cascade sizes for three different values of $\gamma$: 0, 0.045, and 0.06. When $\gamma$ is sufficiently small, all nodes are vulnerable (i.e., their default will trigger the default of all downstream nodes) and the distribution of cascade sizes is exactly the distribution of out-component sizes. Therefore, the results for $\gamma = 0$ represent the distribution of the fraction of nodes that can be reached starting from a randomly chosen node. Interestingly, for $\gamma = 0$ and $\gamma = 0.045$ we see peaks at around $35\%$ for correlated networks in Fig.~\ref{PQ45}(c), but these peaks are absent for edge-uncorrelated networks in Fig.~\ref{PQ45}(d).

For $\gamma=0$, the approximately 0.12 weight at 35\% cascade size in Fig.~\ref{PQ45}(c) is mainly due to Tier-2 seeds. Tier-2 seeds cannot cause the default of Tier-1 banks because of the hierarchical structure of the network (see Fig.~\ref{Network_sketch}), and the resulting cascade size is 35\%, which is the size of the giant component of Tier-2 and Tier-3 subgraph. The 100\% cascades in Fig.~\ref{PQ45}(c) can be triggered exclusively by a Tier-1 seed node because only Tier-1 nodes can have the entire network as their out-component (see Fig.~\ref{Network_sketch}). A Tier-1 node triggers a relatively large number of defaulted edges, that almost certainly results in a 100\% cascade. (With very small probability a Tier-1 seed can result in no cascade, e.g., when it is connected exclusively to Tier-3 nodes, or lead to a 35\% cascade, e.g., when it is connected only to Tier-2 nodes.) There are 10\% Tier-1 nodes in the network, and hence 100\% cascades have probability of approximately $0.1$ in Fig.~\ref{PQ45}(c). Likewise, the main contribution to the 0.78 no-cascade peak in Fig.~\ref{PQ45}(c) is made by Tier-3 (sink) nodes which take 70\% of the network. The extra 0.08 weight to the no-cascade peak is due to Tier-2 nodes which hit exclusively Tier-3 (sink) nodes, and hence fail to trigger a cascade.

Summarizing the above, the three peaks observed in Fig.~\ref{PQ45}(c) appear because of the hierarchical structure of the interbank network, encoded in the edge-correlation matrix $Q$ of Eq.~\eqref{Qjk_mat}. This implies that (i) Tier-3 nodes never trigger a cascade, (ii) only Tier-1 seeds can trigger 100\% cascades, and (iii) within the subnetwork of Tier 2 and 3 nodes, there is a giant component which occupies 35\% of the network; some but not all Tier-2 seeds hit this component.

By comparing the frequency and expected cascade size shown on left and right panels of Fig.~\ref{PQ45}, we see that the edge-correlated interbank structure is more resilient to defaults than the edge-uncorrelated one. This example is of interest to finance, because it shows a new type of robust fragility. Only big banks can bring the entire system down, while medium banks can trigger their subnetwork to collapse. This type of behaviour cannot be observed in edge-uncorrelated models~\cite{GaiKapa10}.

\section{Conclusion}\label{sec:conclusion}
In summary, we have described here an analytical framework which can predict the systemic risk of a networked system of financial institutions. The qualitative type of networks one can address has been extended compared to most existing work, in particular by the inclusion of the non-independent connections between nodes. In this more general setting we find the cascade is described by a vector-valued fixed point problem that reduces to well-understood scalar problems in special cases. We also observed that graph assortativity can strongly affect the course of contagion cascades, and hence showed the importance of incorporating assortativity in numerical and analytical treatments of banking network models. Our analytic framework will enable extensive studies of alternative network topologies. In such studies the cascade condition and cascade frequency provide two easily computed and useful measures of systemic risk by which to compare different network topologies. However, the daunting range of network variables means that both analytical and numerical studies must be carefully framed to address specific issues, for example, to uncover other key determinants of systemic risk. Finally, we anticipate that future work can show how the approach described here may be further extended to include partial recovery models (such as Ref.~\onlinecite{NieYanYorAle07}) and stochastic balance sheets.

\section*{Acknowledgements}
This work was funded in part by Natural Sciences and Engineering Research Council of Canada and the Global Risk Institute for Financial Services of Canada (T.R.H.),  Science Foundation Ireland (11/PI/1026, J.P.G., S.M.), and the Irish Research Council (New Foundations grant, S.M.).

\bibliography{networks}
\end{document}